\documentclass{book}    
\usepackage{piers}  
\usepackage[nolist]{acronym}
\usepackage{amsmath,amssymb,amsfonts}
\usepackage{booktabs}
\usepackage{cite}
\usepackage{xfrac}

\DeclareMathOperator{\sinc}{sinc}

\begin{document}
\begin{acronym}
        \acro{3GPP}{3rd generation partnership project}
        \acro{5G}{fifth generation of mobile networks}
        \acro{6G}{sixth generation of mobile networks}
        \acro{AO}{alternating optimization}
	\acro{AP}{access point}
        \acro{AWGN}{additive white Gaussian noise}
        \acro{DFT}{discrete Fourier transform}
        \acro{LOS}{line-of-sight}
        \acro{MIMO}{multiple-input multiple-output}
        \acro{MISO}{multiple-input single-output}
        \acro{MLP}{multilayer perceptron}
        \acro{NLOS}{non-line-of-sight}
        \acro{NN}{neural network}
        \acro{OFDM}{orthogonal frequency division multiplex}
        \acro{ReLU}{rectified linear unit}
        \acro{RIS}{reconfigurable intelligent surface}
        \acro{SCA}{successive convex approximation}
        \acro{STAR-RIS}{simultaneously transmitting and reflecting reconfigurable intelligent surface}
        \acro{STM}{strongest tap maximization}
        \acro{UE}{user equipment}
\end{acronym}

\title{On the Compromise Between Performance and Efficiency in RIS-aided Communication Systems}
\maketitle

\author      {P. H. C. de Souza}
\affiliation {National Institute of Telecommunications  - Inatel}
\address     {}
\city        {Santa Rita do Sapucaí - MG}
\postalcode  {}
\country     {Brazil}
\phone       {}    
\fax         {}    
\email       {pedro.carneiro@dtel.inatel.br}  
\misc        { }  
\nomakeauthor

\author      {M. Khazaee}
\affiliation {National Institute of Telecommunications  - Inatel}
\address     {}
\city        {Santa Rita do Sapucaí - MG}
\postalcode  {}
\country     {Brazil}
\phone       {}    
\fax         {}    
\email       {masoud@inatel.br}  
\misc        { }  
\nomakeauthor

\author      {L. L. Mendes}
\affiliation {National Institute of Telecommunications  - Inatel}
\address     {}
\city        {Santa Rita do Sapucaí - MG}
\postalcode  {}
\country     {Brazil}
\phone       {}    
\fax         {}    
\email       {luciano@inatel.br}  
\misc        { }  
\nomakeauthor

\begin{authors}

{\bf P. H. C. de Souza}$^{1}$, {\bf M. Khazaee}$^{1}$, {\bf and L. L. Mendes}$^{1}$\\
\medskip
$^{1}$National Institute of Telecommunications  - Inatel, Brazil

\end{authors}

\begin{paper}

\begin{piersabstract}
The \ac{RIS} technology for metasurfaces is ushering in a new paradigm for wireless communication systems. It provides an accessible way for controlling the interaction between electromagnetic waves with the propagation medium. One particularly important aspect is the configuration of the \ac{RIS} elements or reflectors. Simply stated, the objective of the \ac{RIS} configuration is to choose the optimum phase-shift combination that maximizes the channel capacity. Recently, \acp{NN} were proposed for tackling this task and results have shown that the proposed \ac{NN} promotes far less reconfigurations of the \ac{RIS}, consequently reducing the configuration overhead. Beyond that, the \ac{RIS} can be repurposed for tackling the Doppler shift in high-mobility communication systems. Despite not being its usual primary goal, results have also demonstrated that the \ac{RIS} can compensate for the Doppler shift at a small cost in performance. However, the typical reflection-only constraint for \ac{RIS} systems limits the spatial coverage and signal amplification potential achieved by such systems. Therefore, the \ac{STAR-RIS} can be employed to address these limitations by its dual functionality of transmitting and reflecting signals concurrently. It can be shown that the \ac{STAR-RIS} can augment coverage, energy efficiency, and latency reduction, while enhancing sum-rate and physical-layer security across several wireless contexts.
\end{piersabstract}

\psection{Introduction}
\acresetall

Modern wireless communications systems are everchanging with the constant addition of new requirements and technologies alike. The recent advances are not only evident by the ongoing deployment of \ac{5G}, but it becomes even more pronounced in the \ac{6G} \cite{Uusitalo:2021}. Initial requirements for the \ac{6G} include a user-experienced data rate of $1$ Gbit/s, which represents a ten-fold increase relative to the \ac{5G} \cite{akyildiz:20}. One enabling technology that can aid communication systems achieve such goals is the \ac{RIS} \cite{bjorn:22}. 

\acp{RIS} aim are to establish favorable propagation conditions for the transmitted signal. This is done by way of adjustments or configurations on a physical structure placed between the transmitter and receiver, such that the transmitted signal is reflected by this surface with different phase rotations. It can be physically placed in building facades, ceilings, and advertising panels, for example \cite{liaskos:22}. More specifically, the \ac{RIS} is a planar metasurface composed by passive components, such as varactors or varistors with adjustable impedance values \cite{liaskos:22}. Therefore, the \ac{RIS} reflectors act as independent wave scatterers \cite{bjorn:22,liaskos:22}.

Several works analyzed different aspects of \ac{RIS} systems. Great part of them \cite{alexandropoulos:22} are typically concerned only with the algorithms used for configuring the \ac{RIS}, and its performance in terms of channel capacity experienced by the user, for example. While other researchers focus on specific problems surrounding \ac{RIS} systems \cite{basar:21}. Nevertheless, in this work, we strive to present a unified analysis of a \ac{RIS} system considering the last development on phase-shift configuration via \acp{NN} and Doppler shift compensation in high-mobility scenarios. We demonstrate trough numerical results that \acp{NN} are more resource-efficient when configuring the \ac{RIS}. Furthermore, we also show that the Doppler shift can be compensated with minor modifications in well established \ac{RIS} configuration methods. Finally, the \ac{STAR-RIS} systems are also briefly explored in the context of this work.

This work is organized as follows: Section \ref{sec:sysmodel} details the channel model alongside the particularization necessary for the so-called stationary and mobile models; Section \ref{sec:RISconfig} discusses the \ac{RIS} configuration problem and presents the configuration methods studied in this work; Section \ref{sec:num} brings the numerical results and performance analysis of the aforementioned methods; Section \ref{sec:StarRis} gives a brief summary of the perspectives on \ac{STAR-RIS} systems and finally, Section \ref{sec:conclusion} concludes the paper.

\psection{System Model}\label{sec:sysmodel}
For the sake of clarity, the system model description is divided into two parts: (i) Subsection~\ref{subsec:sUser} describes the classical model where the \ac{UE} assumes a fixed position in space for all channel realizations. We refer to this as a stationary \ac{UE} model; and (ii) the model where the \ac{UE} is mobile, as will be further detailed in Subsection~\ref{subsec:mUser}.  

\psubsection{Stationary UE model}\label{subsec:sUser}
Assume that a \ac{RIS} with $N$ reflectors or elements is located in between a transmitting \ac{AP} and a receiving \ac{UE} (both employ single antenna systems), so that the signal transmission can be enhanced by the \ac{RIS}. The operations realized by each \ac{RIS} reflector can be represented mathematically by $\boldsymbol{\omega}_{\theta} = e^{\jmath \boldsymbol{\theta}} \in \mathbb{C}^{N}$, where $\boldsymbol{\theta} = [\theta_0\text{,}\theta_1\text{, }\dots\text{,}\theta_{N - 1}]^\text{T}$. Therefore, each \ac{RIS} reflector is able to apply a continuous phase-shift, that is $\theta_n = \left[-\pi,\pi\right] \ \forall n \in N$, to the impinging signal. Moreover, to maximize the received power and simplify hardware design, we assume $\|\omega_{\theta_n}\|_2^2 = 1$ \cite{jian:22}.

Figure~\ref{fig:sysdiag} depicts the baseline propagation model investigated. It contains a direct channel with $L_d$ paths between the \ac{AP} and \ac{UE}, which are all assumed to be \ac{NLOS} paths. Additionally, there is the composite channel, composed by the cascade of the channels between the \ac{AP} and \ac{RIS} ($L_a$ paths) and \ac{RIS} to the \ac{UE} ($L_b$ paths), which are assumed to be \ac{LOS} dominated channels. The \ac{LOS} channel availability for the $L_a$ and $L_b$ paths is highly likely in this propagation model, due to the advantageous locations that the \ac{RIS} can be deployed \cite{liaskos:22}. However, as can be observed from Figure~\ref{fig:sysdiag}, the direct channel properties can not be altered by the \ac{RIS}. In other words, \ac{RIS} phase-shifts can be only imposed on the signal being propagated through the composite channel. Consequently, for each different path of the composite channel ($L_a L_b$ paths), $N$ new paths are created with a modified phase-shift according to $\theta_n = \left[-\pi,\pi\right] \ \forall n \in N$. As will be further detailed, these phase-shifts are carefully configured in order to establish the coherent combination of signals at the \ac{UE}.
\begin{figure}[h!]
	\centering
	\includegraphics[width=0.5\linewidth,keepaspectratio]{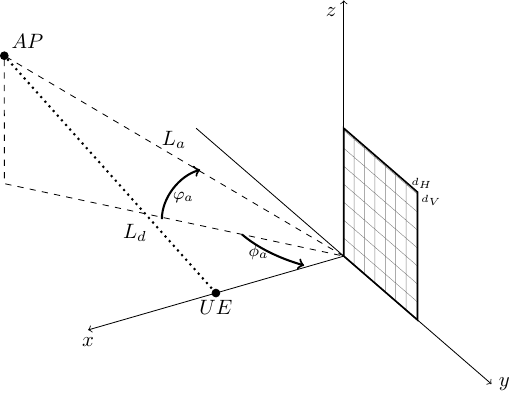}
	\caption{Spatial diagram of the propagation model on a $\left(x\text{, } y\text{, } z\right)$ coordinate system.}\label{fig:sysdiag}
\end{figure}

Also partially illustrated in Figure~\ref{fig:sysdiag} are the relative azimuth angles of arrival ($\phi_a$) and departure ($\phi_b$) at/from the \ac{RIS}, as well as the respective elevation angles $\varphi_a$ and $\varphi_b$. The angles for the $L_b$ channel paths are omitted in Figure~\ref{fig:sysdiag} for convenience, since they have redundant representations. These angles are used to compute the array response of the \ac{RIS} reflectors as a whole. Moreover, note that each reflector has sides of size $d_\text{H}$ and $d_\text{V}$ meters. Consequently, the \ac{RIS} array response considering the $L_a$ channel paths can be given by
\begin{equation}\label{eq:spcsig}
    \mathcal{S}\left(\Phi_a\right) = e^{\jmath {\boldsymbol{\phi}^{(l)}_a}^\text{T} \Psi}\text{, with } \Psi = [\boldsymbol{0}\text{,}\Psi_\text{H}\text{,}\Psi_\text{V}]^\text{T}\text{,}
\end{equation}
where $l \in \{1\text{,}\dots\text{,}L_a\}$ and furthermore we have
\begin{equation}
    \Phi^{(l)}_a = \frac{-2\pi}{\lambda}\left[\cos{\left(\phi^{(l)}_a\right)}\cos{\left(\varphi^{(l)}_a\right)}\text{,}\sin{\left(\phi^{(l)}_a\right)}\cos{\left(\varphi^{(l)}_a\right)}\text{,}\sin{\left(\varphi^{(l)}_a\right)}\right]^\text{T}\text{,}
\end{equation}
in which $\lambda = 3 \times 10^8 f_c^{-1}$, $f_c$ being the central frequency of the signal carrier. Finally, consider the following
\begin{align}
    \Psi_\text{H} &= d_\text{H}\left[\bmod{\left(\sfrac{0}{N_\text{row}}\right)}\text{,}\bmod{(\sfrac{1}{N_\text{row}})}\text{,}\dots\text{,}\bmod{(\sfrac{N-1}{N_\text{row}})}\right]^\text{T}\text{;} \\[.25em]
    \Psi_\text{V} &= d_\text{V}\left[\lfloor\sfrac{0}{N_\text{col}}\rceil \text{,}\lfloor\sfrac{1}{N_\text{col}}\rceil \text{,}\dots\text{,}\lfloor\sfrac{N-1}{N_\text{col}}\rceil\right]^\text{T}\text{,}
\end{align}
for which $N = N_\text{row}N_\text{col}$. Therefore, $\Psi \in \mathbb{C}^{3\times N}$ can be seen as the area spanned by the \ac{RIS} planar surface across the $y-z$ plane, as illustrated in Figure~\ref{fig:sysdiag}. In other words, the \ac{RIS} reflector farthest from the plane origin imposes more rotation to the signal, with the other reflectors presenting gradually less rotation as they draw closer to the plane origin. These rotation values are then weighted by the impinging signal angles, all these operations being computed accordingly in \eqref{eq:spcsig}. The same formulation of \eqref{eq:spcsig} is employed for the $L_b$ channel paths and the \ac{RIS} orientation could be modified without loss of generality.

For the stationary \ac{UE} model, as well as for the mobile \ac{UE} model, we consider that the the \ac{OFDM} system is employed for signal transmission. Therefore, let the unmodulated signal of the composite channel be written as in:
\begin{equation}\label{eq:rxsignal}
    \check{h}\left[m\right] = \sum_{n=0}^{N-1}\sum_{l=1}^{L_a}\sum_{\ell=1}^{L_b}\sqrt{\alpha^{(l)} \beta^{(\ell)}}\mathcal{S}\left(\Phi_a\right)\mathcal{S}\left(\Phi_b\right) \times e^{-\jmath 2\pi f_c\left(\tau^{(l)}_a + \tau^{(\ell)}_b + \tau_{\theta_n}\right)} s\left(m\right)\text{;}
\end{equation}
in which,
\begin{equation}
    s\left(m\right) = \sinc\left(m + B\left(\eta - \tau^{(l)}_a - \tau^{(\ell)}_b + \tau_{\theta_n}\right)\right)\text{,}
\end{equation}
and where $\alpha^{(l)}$ and $\beta^{(\ell)}$ are, respectively, the propagation losses for the $L_a$ and $L_b$ channels, $\mathcal{S}\left(\Phi_a\right)$ and $\mathcal{S}\left(\Phi_b\right)$ are given by \eqref{eq:spcsig}, $\tau^{(l)}_a$ and $\tau^{(\ell)}_b$ are the propagation delays, whereas ${\tau_{\theta_n} = \theta_n / 2\pi f_c}$ is the phase-shift caused by the $n$-th \ac{RIS} element. Furthermore, $B$ is the bandwidth occupied by all subcarriers and $\eta = \min{(\tau^{(l)}_d)}$, $\forall l \in L_d$, to assure causality. Likewise, the direct channel then can be given as
\begin{equation}\label{eq:signal}
    \bar{h}\left[m\right]  = \sum_{l=1}^{L_d}\sqrt{\delta^{(l)}} \sinc\left(m + B \left(\eta - \tau^{(l)}_d\right)\right) \times e^{-\jmath 2\pi \left(f_c \tau^{(l)}_d\right)} \nonumber\text{,}
\end{equation}
wherein $\delta^{(l)} \ \forall l \in L_d$ are the propagation losses for the $L_d$ direct channel paths.

In the next subsection we present the mobile \ac{UE} model which is similar to the one presented in the current subsection, but where key modifications are introduced to properly represent a non-stationary \ac{UE}. As will become clear, the mobile \ac{UE} model is employed in this work for the investigation of the Doppler effect in \ac{RIS} systems.

\psubsection{Mobile UE model}\label{subsec:mUser}
For the mobile \ac{UE} model it is assumed that the \ac{UE} moves at a speed of $v$ m/s, while receiving the signal transmitted by a stationary \ac{AP} located far from the \ac{UE} trajectory. Similarly to Subsection~\ref{subsec:sUser}, let a stationary \ac{RIS} of $N$ reflectors be available to assist communications between the \ac{AP} and the mobile \ac{UE}. The propagation model also remains practically the same to the baseline illustrated in Figure~\ref{fig:sysdiag}. However, note that by considering roadside \acp{RIS} deployment, for example, then we can assure the availability of \ac{LOS} dominated channels for the composite channel \cite{wu:21}. Therefore, Figure~\ref{fig:sysDdiag} illustrates the discussed mobile \ac{UE} model, where multiple roadside \acp{RIS} provide seamless signal coverage to the mobile \ac{UE}. It is also important to note that we now assume the transmission of signal frames, each frame consisting of $U$ transmitted blocks (see Figure~\ref{fig:sysDdiag}). Nevertheless, a constant channel for the duration of a transmitted frame is considered. Since the distance traveled by the \ac{UE} is negligible within the frame duration\footnote{The frame duration is typically in the order of milliseconds (see Section~\ref{sec:num}).}, we also consider that the \ac{UE} speed, $v$, is approximately constant. Moreover, for the sake of simplicity, we assume that the channel coefficients and the Doppler frequency estimations are ideal, which can be achievable in practice (near ideal) via estimation methods such as those demonstrated in \cite{wu:21}. In conclusion, see that our focus is to optimize the phase-shifts configuration for a single \ac{RIS} within each frame transmission.
\begin{figure}[!h]
	\centering
	\includegraphics[width=0.5\linewidth,keepaspectratio]{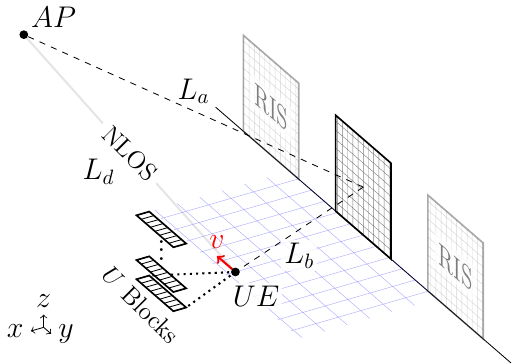}
	\caption{Diagram of the \ac{RIS}-aided mobile \ac{UE} model presented in this work. Note that the \ac{RIS} size and distances dimensions are not to scale.}\label{fig:sysDdiag}
\end{figure}

Differently form the stationary \ac{UE} model (see Subsection~\ref{subsec:sUser}), recall that we now consider $K$ subcarriers transmitting $U$ data blocks using an \ac{OFDM} system. Therefore, the unmodulated signal transmitted via the composite channel can be described by
\begin{equation}\label{eq:rxDsignal}
    \check{H}_{u,m} = \sum_{n=0}^{N-1}\sum_{l=1}^{L_a}\sum_{\ell=1}^{L_b}\sqrt{\alpha^{(l)} \beta^{(\ell)}}\mathcal{S}\left(\Phi_a\right)\mathcal{S}\left(\Phi_b\right) s\left(m\right)\times e^{-\jmath 2\pi \left(f_c\left(\tau^{(l)}_a + \tau^{(\ell)}_b + \tau_{\theta_{u,n}}\right) - \check{f}^{(\ell)}_D uT_{\text{b}}\right)}\text{;}
\end{equation}
in which,
\begin{equation}
    s\left(m\right) = \sinc\left(m + B \left(\eta - \tau^{(l)}_a - \tau^{(\ell)}_b + \tau_{\theta_{u,n}}\right)\right)\text{,}
\end{equation}
where $\check{f}^{(\ell)}_D$ is the Doppler frequency due to the motion of the \ac{UE} relative to the stationary \ac{RIS} reflectors, $f_c$ defines the central frequency of the signal carrier, $u \in \{1\text{,}2\text{,}\ldots\text{,}U\}$ is the transmission block index, $T_{\text{b}}$ is the block duration in seconds and $B$ is the total bandwidth. Additionally, note that ${\tau_{\theta_{u,n}} = \theta_{u,n} / 2\pi f_c}$ is the phase-shift configured for the $n$-th \ac{RIS} element during the $u$-th block transmission. It follows then that the direct channel can be defined as
\begin{equation}\label{eq:dsignal}
    \bar{H}_{u,m} = \sum_{l=1}^{L_d}\sqrt{\delta^{(l)}} \sinc\left(m + B \left(\eta - \tau^{(l)}_d\right)\right)\times e^{-\jmath 2\pi \left(f_c \tau^{(l)}_d - \bar{f}_D uT_{\text{b}}\right)} \nonumber\text{,}
\end{equation}
wherein $\bar{f}_D$ is the Doppler frequency due to the motion of the \ac{UE} relative to the \ac{AP}. Finally, observe that the mobile \ac{UE} model specializes to the stationary model when $v = 0$ m/s (hence $\check{f}^{(\ell)}_D = 0 \text{ Hz} \ \forall \ell$ and $\bar{f}_D = 0$ Hz) and $U = 1$, as expected. 

\psection{RIS Configuration}\label{sec:RISconfig}
There is no unique phase-shift configuration for the \ac{RIS} that can simultaneously maximize the channel capacity at the \ac{UE} for all $K$ subcarriers \cite{bjorn:22}. This so because maximizing the channel capacity demands that the direct and composite channels combine constructively or coherently at the \ac{UE}, with their respective phases aligned. The problem is that each subcarrier presents different channel properties as, for example, phase-shifts. Therefore, if we assume a constant \ac{RIS} configuration for a given bandwidth, it becomes unfeasible to configure a different $\boldsymbol{\omega}_{\theta}$ for each subcarrier. Consequently, a trade-off must be established considering all subcarriers, so that the best \ac{RIS} configuration can be determined. Moreover, also recall that the \ac{RIS} can be leveraged to compensate the frequency shift cause by the Doppler effect. Nevertheless, to further define the \ac{RIS} configuration problem at hand, let us begin with the achievable rate definition:
\begin{equation}\label{eq:rate}
    R = \frac{B}{\xi}\sum_{u=1}^{U}{\sum_{i=0}^{K - 1}{\log_2{\left(1 + \frac{p^{(u)}_i\|\mathbf{f}_i^\text{H} \mathbf{\bar{h}}_u + \mathbf{f}_i^\text{H} \mathbf{V}_u^{\text{T}}\boldsymbol{\omega}_{\theta_u}\|_2^2}{BN_0}\right)}}} \ \sfrac{\text{bit}}{\text{s}}\text{,}
\end{equation}
wherein 
\begin{align}
        &\mathbf{\bar{h}}_u = \left[\bar{h}\left[0\right]\text{,}\bar{h}\left[1\right]\text{,}\dots\text{,}\bar{h}\left[M-1\right]\text{,}\mathbf{0}_{K - M}\right]^\text{T} \text{(Stationary \ac{UE} model)}\text{;} \nonumber \\[0.5em]
        &\mathbf{\bar{h}}_u = \left[\bar{H}_{u,0}\text{,}\bar{H}_{u,1}\text{,}\dots\text{,}\bar{H}_{u,M-1}\text{,}\mathbf{0}_{K - M}\right]^\text{T} \text{(Mobile \ac{UE} model)}\text{.} \nonumber
\end{align}
Furthermore, we have
\begin{align}
        &\left[\check{h}\left[0\right]\text{,}\check{h}\left[1\right]\text{,}\dots\text{,}\check{h}\left[M-1\right]\text{,}\mathbf{0}_{K - M}\right]^\text{T} = \mathbf{V}_u^{\text{T}} \boldsymbol{\omega}_{\theta_u} \text{ (Stationary \ac{UE} model)}\text{;} \nonumber \\[0.5em]
        &\left[\check{H}_{u,0}\text{,}\check{H}_{u,1}\text{,}\dots\text{,}\check{H}_{u,M-1}\text{,}\mathbf{0}_{K - M}\right]^\text{T} = \mathbf{V}_u^{\text{T}} \boldsymbol{\omega}_{\theta_u}\text{ (Mobile \ac{UE} model)}\text{.} \nonumber
\end{align}
Notice that in \eqref{eq:rate} $M$ represents the total number of signal samples, $\xi = K + M - 1$ accounts for the cyclic prefix loss, $\mathbf{f}_i \in \mathbb{C}^{K}$ represents the $i$-th row of the \ac{DFT} matrix $F_{i,j} = e^{-\jmath 2\pi ij/K}$, $\mathbf{p} \in \mathbb{R}^{K}$ is the power vector, with $p^{(u)}_i$ being the power allocated to the $k$-th subcarrier at the $u$-th block\footnote{We assume that the transmitted blocks are mutually independent.}, and $N_0$ is the \ac{AWGN} power density. More specifically, for the stationary \ac{UE} model, \eqref{eq:rate} specializes with $U = 1$, as stated before, accompanied by the appropriate coefficients representations as described above.

\psubsection{Alternating Optimization}
Considering the achievable rate defined in \eqref{eq:rate} as the actual channel capacity at the \ac{UE}, let us then cast the capacity maximization problem as an optimization problem. To perform the \ac{RIS} configuration, one can solve successive convex problems \cite{yang:20}. For this, let us also introduce a set of auxiliary variables $y_i \ \forall i$, $a_i \ \forall i$, $b_i \ \forall i$, and define:
\begin{equation}
    f_i\left(a_i,b_i\right) \triangleq \Tilde{a}_i^2 + \Tilde{b}_i^2 + 2\Tilde{a}_i(a_i - \Tilde{a}_i) + 2\Tilde{b}_i(b_i - \Tilde{b}_i);
\end{equation}
which give us the following optimization problem (Stationary \ac{UE} model)
\begin{alignat}{3}
&\underset{\boldsymbol{\omega}_{\theta}\text{,}\{y_i\}\text{,}\{a_i\}\text{,}\{b_i\}}{\text{maximize }} &&\sum_{i=0}^{K - 1}{\log_2{\left(1 + \frac{p_iy_i}{BN_0}\right)}} \nonumber \\
&\text{subject to } &&\|\omega_{\theta_{u,n}}\|_2^2 \leq 1 \ , \ &&\forall n \ , \nonumber \\
                    & &&a_i = \Re\{\mathbf{f}_i^\text{H} \mathbf{\bar{h}}_u + \mathbf{f}_i^\text{H} \mathbf{V}_u^{\text{T}}\boldsymbol{\omega}_{\theta_u}\} \ , \ &&\forall i \ , \label{eq:cvxprob}\\
                    & &&b_i = \Im\{\mathbf{f}_i^\text{H} \mathbf{\bar{h}}_u + \mathbf{f}_i^\text{H} \mathbf{V}_u^{\text{T}}\boldsymbol{\omega}_{\theta_u}\} \ , \ &&\forall i \ , \nonumber \\
                    & &&y_i \leq f_i\left(a_i,b_i\right) \ , \ &&\forall i \ , \nonumber
\end{alignat}
wherein the power allocation, $\mathbf{p}$, and the \ac{RIS} phase-shifts, $\boldsymbol{\omega}_{\theta}$, are alternately optimized. More precisely, a solution for \eqref{eq:cvxprob} can be found by successively updating $\{\Tilde{a}_i\}$ and $\{\Tilde{b}_i\}$. Hence the reason why this \ac{RIS} configuration method is denoted as \ac{AO}. Note that in this work we use the \ac{AO} as a benchmark for numerical analysis purposes, and thus we do not delve deeper into more technical features regarding it. For interested readers, more details about the \ac{AO} convergence guarantees and other definitions can be found in \cite{yang:20}. 

\psubsection{Strongest Tap Maximization}
Finding a solution (\ac{RIS} configuration) for \eqref{eq:cvxprob} may entail a prohibitive computational complexity \cite{yang:20}. However, configuring the \ac{RIS} using the time domain as a reference, may be more straightforward than finding an optimum trade-off for $\boldsymbol{\omega}_{\theta_u}$ considering all $K$ subcarriers. Specially when realizing that usually $M \ll K$ and also that for \ac{LOS} channels most of the received signal power is concentrated in a few time-domain samples \cite{bjorn:22}. Therefore, we have the so-called \ac{STM} method for computing the \ac{RIS} configuration, which is given by
\begin{align}
    \boldsymbol{\omega}_{\theta_u}^{(m^\ast)} &= \underset{m \in \left\{0,1,\dots,M-1\right\}}{\arg \max}{\|\bar{h}_u[m] + \left[\mathbf{V}_u\right]_m^\text{T} \boldsymbol{\omega}_{\theta_u}^{(m)} \|_2^2}\text{, wherein}\label{eq:tvStm} \\[.5em]
    \boldsymbol{\omega}_{\theta_u}^{(m)} &= e^{\jmath \left(\arg{\left\{\bar{h}_u\left[m\right]\right\}} - \arg{\left\{\left[\mathbf{V}_u\right]_m\right\}}\right)} \ , \ \forall u \in \{1\text{,}2\text{,}\ldots\text{,}U\}\text{.} \nonumber
\end{align}
The \ac{STM} method computes the suboptimal \ac{RIS} configuration, $\boldsymbol{\omega}_{\theta}^{(m^\ast)}$, that results in the largest magnitude of the received time-domain signal. Nonetheless, observe that for the mobile \ac{UE} model the \ac{STM} requires a time-varying \ac{RIS} that is capable of configuring a new phase-shift for each new transmitted block. Therefore, we henceforth denote the \ac{STM} of \eqref{eq:tvStm} as the time-varying \ac{STM} (TV-\ac{STM}).

\psubsubsection{Time-invariant Strongest Tap Maximization}
In the context of the mobile \ac{UE} model, the Doppler effect will essentially introduce signal distortions in \eqref{eq:rxsignal} and \eqref{eq:dsignal}, but these signals still present a significant degree of similarity across different transmission blocks. With that in mind, one may conceive a time-invariant \ac{RIS} configuration, in order to leverage such similarity \cite{pedro:24}. The time-invariant configuration not only simplifies the \ac{RIS} operation, but can also compensate the frequency shift caused by the Doppler effect. This is clarified further by the numerical results available in this work (see Section \ref{sec:num}). Therefore, let the time-invariant \ac{STM} (TI-\ac{STM}) be given by
\begin{align}
    \boldsymbol{\omega}_{\theta_{u^{\prime}}}^{(m^\ast)} &= \underset{m \in \left\{0,1,\dots,M-1\right\}}{\arg \max}{\|\bar{h}_u[m] + \left[\mathbf{V}_u\right]_m^\text{T} \boldsymbol{\omega}_{\theta_{u^{\prime}}}^{(m)} \|_2^2}\text{, wherein} \label{eq:tiStm} \\[.5em]
    \boldsymbol{\omega}_{\theta_{u^{\prime}}}^{(m)} &= e^{\jmath \left(\arg{\left\{\bar{h}_{u^{\prime}}\left[m\right]\right\}} - \arg{\left\{\left[\mathbf{V}_{u^{\prime}}\right]_m\right\}}\right)} \ , \ u^{\prime} \in \{1\text{,}2\text{,}\ldots\text{,}U\}\text{.} \nonumber
\end{align}
The time-invariant \ac{STM} configuration is realized by choosing a fixed transmission block index, $u^{\prime}$, such that a reference phase-shift is established. Since the blocks of information are transmitted sequentially, we thus adopt $u^{\prime} = 1$ throughout this work as the fixed transmission block index. Keep in mind the time-invariant \ac{STM} configuration does not promote the best phase alignment between direct and composite channels, but the reduced frequency shift and operational complexity compensate for that.

\psubsection{Neural Network}
A neural network architecture was proposed in \cite{pedro:25} for configuring the \ac{RIS}, its modeling being mostly derived from the formulations provided by \eqref{eq:cvxprob} and \eqref{eq:tvStm}. Therefore, let the \ac{NN} architecture consists of $\ell \in \{0\text{,} 1\text{,} \ldots\text{,} L\}$ layers, with the input layer ($\ell = 0$) computing the following string of operations: 
\begin{equation}\label{eq:inputlayer}
    \boldsymbol{\theta}_0 = \text{ReLU}\left(\sum_{n}{\left[\boldsymbol{\Theta}\mathbf{W}_0\right]_n}\right) + \mathbf{b}_0\text{;}
\end{equation}
where $\sum_{n}\left[\cdot\right]_n$ is the sum of all columns of the resulting matrix within the  argument, and in which
\begin{equation}
    \boldsymbol{\Theta} = \mathbf{1}_N \arg{\left\{\mathbf{\bar{h}}_u^{\text{T}}\mathbf{F}^{\text{H}}\right\}} - \arg{\left\{\mathbf{V}_u\mathbf{F}^{\text{H}}\right\}} \, \in \mathbb{R}^{N \times K} \text{,}
\end{equation}
represents the phase-shift alignment for all subcarriers and \ac{RIS} elements. Moreover, note that
\begin{equation}
    \text{ReLU}\left(z\left[i\right]\right) = \max\left(0,z\left[i\right]\right)\text{,} \ \forall i\text{,}
\end{equation}
is the so-called \ac{ReLU} function, and both $\mathbf{W}_0 \in \mathbb{R}^{K \times N}$ and $\mathbf{b}_{0} \in \mathbb{R}^{N}$ are the \ac{NN} learnable parameters. Finally, also see that    
\begin{equation}\label{eq:layers}
    \boldsymbol{\theta}_{\ell} = \text{ReLU}\left(\boldsymbol{\theta}_{\ell - 1} \odot \mathbf{w}_{\ell}\right) + \mathbf{b}_{\ell} \ \text{,} \ \ell \in \left\{1,\dots,L\right\}\text{,}
\end{equation}
for which $\mathbf{w}_{\ell}\text{, }\mathbf{b}_{\ell} \in \mathbb{R}^{N}$ and $\boldsymbol{\omega}_{\theta_L} = e^{\jmath \boldsymbol{\theta}_L}$ represents the configured phases by the \ac{NN}.

Notice that entries of $\mathbf{W}_0$ in \eqref{eq:inputlayer} are weighted on the non-aligned phases between the direct channel, $\bar{h}_u$, and the composite channels, $\mathbf{V}_u$. During the \ac{NN} training, these entries are modified in order to achieve the best \ac{RIS} configuration trade-off, taking into account all $K$ subcarriers and $N$ \ac{RIS} elements. In other words, these are the learnable parameters that are optimized at the training stage and empower the \ac{NN} to learn a task of interest, this being the \ac{RIS} configuration. The interested reader can find more information in \cite{pedro:25} about this \ac{NN} in particular and also in \cite{zapp:19} for more general discussions about \acp{NN}. 

\psection{Numerical Results}\label{sec:num}
In this section, computer simulations are employed for generating numerical results regarding the \ac{RIS} configuration methods discussed in Section~\ref{sec:sysmodel}. Although the achievable rate expressed in \eqref{eq:rate} is used as a baseline performance metric for comparing the different methods, we also analyze the efficiency rate. In general, this metric quantifies the overhead or the amount of resources used for configuring the \ac{RIS}. In particular, the efficiency rate grows as fewer \ac{RIS} reflectors need to change their phase configuration between adjacent channel realizations \cite{pedro:25}. For computing the efficiency rate, we assume a quantized \ac{RIS} configuration that can be written as
\begin{equation}\label{eq:quantzlvl}
    \boldsymbol{\theta}^q = \left\lfloor\frac{\boldsymbol{\theta}}{\Delta}\right\rceil\times\Delta;\text{ in which }\Delta = \frac{\pi}{2^{\left(b - 1\right)}} \text{,}
\end{equation}
wherein $\boldsymbol{\omega}_{\theta^q} = e^{\jmath \boldsymbol{\theta}^q}$ describes the discretized \ac{RIS} configuration and the number of bits or levels of quantization is given by $b$-bit. More specifically, the efficiency rate calculation is based on the number of \ac{RIS} reflectors that require phase-shift reconfiguration over the total number of reflectors and channel realizations. The remaining relevant system parameters and the \ac{NN} specifications are all defined according to guidelines found in \cite{pedro:24,pedro:25}; we refer the interested readers to these works for more information.

In Figure~\ref{fig:freqplot} we have the achievable rate of various \ac{RIS} configuration methods discussed in this work, traced for a range of bandwidth values, $B$, and also assuming $N = 400$ reflectors for the \ac{RIS}. Moreover, notice in Figure~\ref{fig:freqplot} that the coherent rate is also illustrated: it represents the ideal scenario whereby the direct channel coefficients sum coherently with the composite channel counterparts. The coherent rate can be seen as an upper bound to the achievable rate \cite{bjorn:22,pedro:25}. Therefore, observe in Figure~\ref{fig:freqplot} that the achievable rate increases practically linearly with the bandwidth. Furthermore, see that the performance of all methods are virtually identical and that they follow closely the coherent rate values.
\begin{figure}[h!]
    \centering
    \includegraphics[width=0.5\linewidth,keepaspectratio]{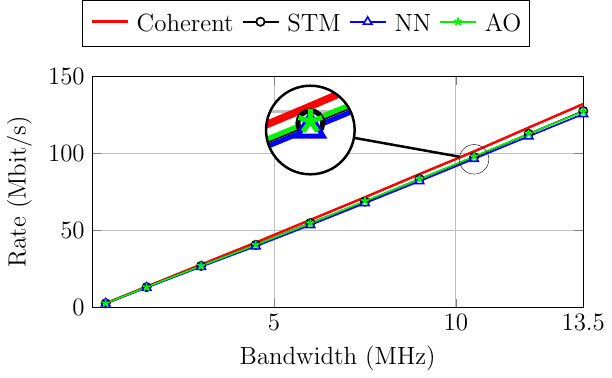}
    \caption{Achievable rate for multiple bandwidth values.}\label{fig:freqplot}
\end{figure}

With the results of Figure~\ref{fig:freqplot} in mind, we bring in Table~\ref{tbl:eff} the results for efficiency rate. For Table~\ref{tbl:eff}, the value of $B = 10.5$ MHz is fixed, while different values for $N$ are presented and a $4$-bit quantization is assumed for the \ac{RIS} configuration\footnote{The performance obtained with $4$-bit quantization is equivalent to that achieved by the continuous \ac{RIS} configuration \cite{pedro:25}.}. Note in Table~\ref{tbl:eff} that the \ac{NN} is significantly more efficient than other discussed configuration methods, regardless of the number of \ac{RIS} reflectors considered. In summary, the phase-shifts reconfigurations are remarkably rare for the \ac{NN}. 
\begin{table}[h!]
    \caption{Efficiency rate for a $4$-bit quantization of the RIS configuration.}\label{tbl:eff}
        \centering
	\setlength{\tabcolsep}{1mm}
	\resizebox{0.8\linewidth}{!}{%
		\begin{tabular}{p{0.15\linewidth}p{0.25\linewidth}p{0.25\linewidth}p{0.25\linewidth}} \toprule 
                \textbf{Number of} & & \textbf{Configuration Methods} & \\
                \cmidrule(lr){2-4} 
                \textbf{Reflectors} & STM (4-bits) & AO (4-bits) & NN (4-bits) \\
                \cmidrule(lr){1-1}\cmidrule(lr){2-2}\cmidrule(lr){3-3}\cmidrule(lr){4-4}
                $N = 100$ & $12 \%$ & $14.2 \%$ & $86.2 \%$ \\[.5em]
                $N = 196$ & $12.2 \%$ & $17.5 \%$ & $91.5 \%$ \\[.5em]
			$N = 324$ & $12.4 \%$ & $19 \%$ & $94 \%$ \\ \bottomrule
	\end{tabular}}
\end{table}

The numerical results discussed thus far in this section were in the context of the stationary \ac{UE} model. However, in what follows we show results obtained for the TV-\ac{STM} of \eqref{eq:tvStm} and the TI-\ac{STM} of \eqref{eq:tiStm}, in the context of the mobile \ac{UE} model. Recall that the main objective of the mobile \ac{UE} model is to foment the investigation of the Doppler effect in \ac{RIS} systems, such as the one discussed in this work (see Section~\ref{sec:sysmodel}). Therefore, in Table~\ref{tbl:doppler} we have performance results in terms of achievable rate, considering again a fixed bandwidth of $B = 10.5$ MHz for different \ac{RIS} sizes. As demonstrated in \cite{pedro:24}, the TI-\ac{STM} configuration is able to fully compensate the Doppler shift in the frequency ($f_D \sim 0$ Hz) of the signal received at the \ac{UE}. Nevertheless, in Table~\ref{tbl:doppler} is possible to verify that as the number of \ac{RIS} reflectors increases, the smaller is the drop of performance of the TI-\ac{STM} configuration in relation to the TV-\ac{STM}. Consequently, through the aid of Table~\ref{tbl:doppler}, it becomes clear that the difference in performance due to the Doppler shift compensation can be considered negligible for large \acp{RIS}.
\begin{table}[h!]
    \caption{Impact on the rate performance due to the Doppler shift compensation.}\label{tbl:doppler}
        \centering
	\setlength{\tabcolsep}{1mm}
	\resizebox{0.8\linewidth}{!}{%
		\begin{tabular}{p{0.15\linewidth}p{0.15\linewidth}p{0.25\linewidth}p{0.25\linewidth}} \toprule 
                \textbf{Number of} & & \textbf{Configuration Methods} & \\
                \cmidrule(lr){2-4} 
                \textbf{Reflectors} & Coherent & TV-STM & TI-STM ($f_D \sim 0$ Hz) \\
                \cmidrule(lr){1-1}\cmidrule(lr){2-2}\cmidrule(lr){3-3}\cmidrule(lr){4-4}
                $N = 25$ & $47.8$ Mbit/s & $40.6$ Mbit/s ($\downarrow 15 \%$) & $38$ Mbit/s ($\downarrow 20.5 \%$) \\[.5em]
                $N = 100$ & $68.6$ Mbit/s & $61.3$ Mbit/s ($\downarrow 10.6 \%$) & $56.9$ Mbit/s ($\downarrow 17 \%$)\\[.5em]
			$N = 400$ & $101.3$ Mbit/s & $97.8$ Mbit/s ($\downarrow 3.4 \%$) & $96.1$ Mbit/s ($\downarrow 5.1 \%$) \\ \bottomrule
	\end{tabular}}
\end{table}

The variety of results unveiled in this section shows that \ac{RIS} systems should be investigated taking into account its different characteristics. By establishing a baseline performance (e.g. achievable rate), one can then evaluate different \ac{RIS} functionalities (e.g. configuration efficiency and/or Doppler effects) and their impact on such performance.       

\psection{Perspectives on STAR-RIS Systems}\label{sec:StarRis}
In a traditional \ac{RIS} configuration, as we have discussed in this work, it is necessary for the transmitter and receiver to be located on the same side. This is because the \ac{RIS} is specifically intended to reflect wireless signals back in the direction from which they originated. If there are users on both sides of a \ac{RIS}, then severe limitations on its performance are to be expected. In order to tackle this difficulty, novel solutions have been suggested that make use of \ac{STAR-RIS} systems, by leveraging the simultaneous reflection and transmission of signals \cite{liu:21}.

It is envisioned that \ac{STAR-RIS} systems may function on the basis of ``on/off'' (transmission or reflection) protocols. This can be seen as a combination of a conventional \ac{RIS} that solely reflects or transmits signals at each time. Alternatively, as a more sophisticated approach, one \ac{STAR-RIS} system protocol may involve the simultaneous reflection and transmission of an element of the incoming electromagnetic waves. The surface divides the incoming signal into two components, with one component being reflected towards users on the same side, while the other is directed to users located on the opposite side. The fundamental advantage of this protocol being its ability to simultaneously support numerous users on both sides of the surface. Other \ac{STAR-RIS} protocol might even be based on the switching of all elements between transmission and reflection modes at orthogonal time intervals.

The \ac{STAR-RIS} performance can be analyzed in a variety contexts, such as: coverage enhancement, physical layer security improvement, latency reduction, sum rate enhancement (see Section~\ref{sec:RISconfig}), energy efficiency enhancement, interference mitigation and resource allocation. These are essential metrics for evaluating the capability of \ac{STAR-RIS} systems and its perspectives within the context of next-generation networks.

\psection{Conclusion}\label{sec:conclusion}
In this work, we offered a unified analysis of different aspects of \ac{RIS} systems. For this, we first introduced the stationary/mobile \ac{UE} model for the \ac{RIS} system. This allowed the investigation of the channel capacity maximization at the \ac{UE}, as well as the Doppler shift compensation, both aided by the \ac{RIS}. In particular, it was demonstrated via numerical results that the \ac{NN}-based configuration for the \ac{RIS} is more resource-efficient than other well established configuration methods. Moreover, we highlighted the fact that the Doppler shift can indeed be compensated with the aid of \ac{RIS} systems at a negligible cost in performance.

\ack
This work was partially funded by project XGM-AFCCT-2024-2-15-1 funded by xGMobile – EMBRAPII Inatel Competence Center on 5G and 6G Networks, with financial resources from the PPI IoT/Manufatura 4.0 from MCTI grant number 052/2023, signed with EMBRAPII; by the Samurai Project funded by FAPESP (Grant No. 20/05127-2); by FAPEMIG under the contract No PPE-00124-23; and by CNPq-Brasil.








\end{paper}

\end{document}